\documentclass{article}

\usepackage[nonatbib]{neurips_2020}

\usepackage[utf8]{inputenc} 
\usepackage[T1]{fontenc}    
\usepackage{hyperref}       
\usepackage{url}            
\usepackage{booktabs}       
\usepackage{amsfonts}       
\usepackage{nicefrac}       
\usepackage{microtype}      
\usepackage{graphicx}
\usepackage{wrapfig}

\usepackage{floatrow}
\newfloatcommand{capbtabbox}{table}[][\FBwidth]

\title{To be a fast adaptive learner: using game history to defeat opponents}

\author{%
  David S.~Hippocampus\thanks{Use footnote for providing further 
    about author (webpage, alternative address)---\emph{not} for 
    funding agencies.} \\
  Department of Computer Science\\
 Cranberry-Lemon University\\
 Pittsburgh, PA 15213 \\
 \texttt{hippo@cs.cranberry-lemon.edu} \\
}

\begin{document}

\maketitle

\begin{abstract}
  In many real-world games, such as trader repeatedly bargaining with customers, it is very hard for a single AI trader to make good deals with various customers in a few turns, since customers may adopt different strategies even the strategies they choose are quite simple. In this paper, we model this problem as fast adaptive learning in the finitely repeated games. We believe that past game history plays a vital role in such a learning procedure, and therefore we propose a novel framework (named, F3) to fuse the past and current game history with an Opponent Action Estimator (OAE) module that uses past game history to estimate the opponent’s future behaviors.  The experiments show that the agent trained by F3 can quickly defeat opponents who adopt unknown new strategies. The F3 trained agent obtains more rewards in a fixed number of turns than the agents that are trained by deep reinforcement learning. Further studies show that the OAE module in F3 contains meta-knowledge that can even be transferred across different games.
\end{abstract}

\section{Introduction}

Game Theory can be defined as the study of mathematical models of conflict and cooperation between intelligent rational decision-makers.~\cite{ref1} It can be applied in different ambit of Artificial Intelligence, such as AI Pluribus in multiplayer poker~\cite{ref2, ref3},  Multi-agent AI systems~\cite{ref4, ref5}, Imitation and Reinforcement Learning~\cite{ref6, ref7} and so on.

Interactions in the real world, like negotiation, are challenging tasks. The opponents are usually various and the turns of interactions are usually multiple but limited. However previous interactions, historical information, are rarely used. This kind of interaction can be modeled as a finitely repeated game. In the face of various opponents or strategies, how to make the agent adapt quickly is an important issue. Many new strategies are integration or variants of the old so we can use historical information to make the agent have fast adaptability.

The repeated game consists of a number of repetitions of some base game (also called a stage game)~\cite{ref8} which may be broadly divided into two classes, finite and infinite, depending on how long the game is being played for. Finite games are those in which both players know that the game is being played a specific and finite number of rounds. When the game ends, all the players will receive end terminal~\cite{ref9} and final rewards.

In repeated games, game history is the main basis for making decisions and defeating opponents. Memory is a special form of knowledge~\cite{ref9}. Judging the type of opponent, inferring incomplete information, and predicting the behavior of the opponent all depend on it. In the absence of other additional information from the opponent, historical information can be said to be the sole basis.

\begin{figure}
  \centering
  \label{history}
  \includegraphics[width=12cm]{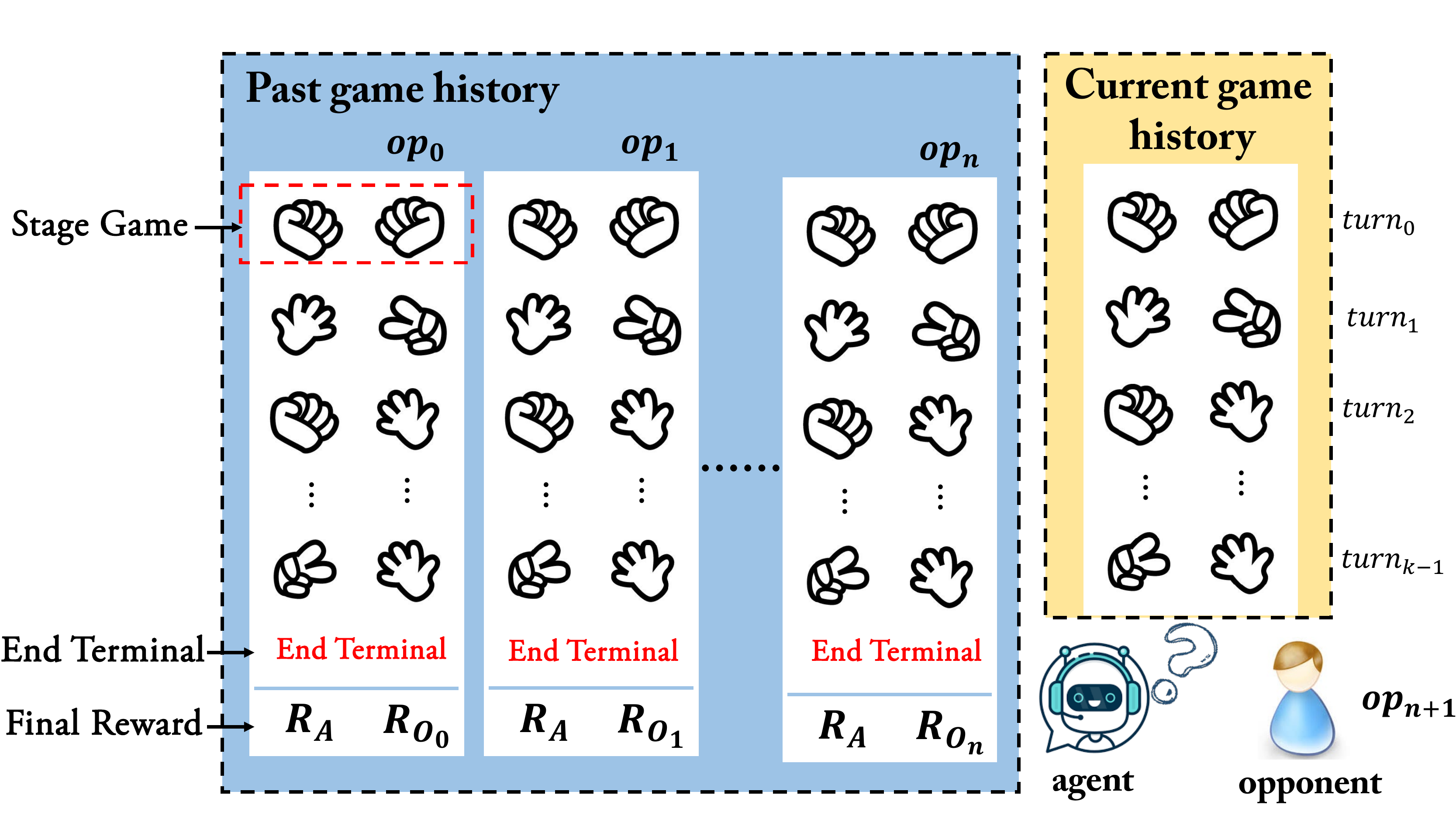} 
  \caption{Past game history and Current game history.}
\end{figure}

But historical information has different types and different history have different meanings. The set of histories which have terminals~\cite{ref9} can be defined as \textbf{\emph{Past game history}} and the set of actions that have been taken in the current event can be defined as \textbf{\emph{Current game history}}. \emph{Current game  history} consists of histories after which the chance player and the decision maker move~\cite{ref9}. As shown in Figure~\ref{history}, in finitely repeated games, the \emph{Past game history} is the set of events that have an end terminals and have received the final rewards. The \emph{Current game history} is the set of actions that have already made in the ongoing game.

Current studies about the use of historical information in repeated games are mainly limited to the \emph{Current game history} (e.g.~\cite{ref9, ref10, ref11}). The studies about modeling opponents (e.g.~\cite{ref12, ref13}) treat all their opponents as one type. But in the real world, the type of opponents may be fickle. In a new game, we may meet opponents that we have never encountered before. Different from one-shot interactions, players would consider more about the long-term payoffs.

In finite repeated games, the opponents' many new strategies are integration or variants of the old. For example, in Iterated Prisoners Dilemma (IPD), \emph{Spiteful Tit For Tat} (Prison1998), \emph{Hard Tit For Two Tats}~\cite{ref14} and \emph{Suspicious Tit For Tat} ~\cite{ref15} are all proposed based on the \emph{Tit For Tat}~\cite{ref16}. The \emph{Past game history} information library stores the strategies that have been encountered and the best response under various circumstances. When agents encounter a new strategy, they can infer or learn new strategies quickly to defeat their opponents from the similar strategies in the \emph{Past game history}. 

\begin{figure}
\centering 
\includegraphics[width=14cm]{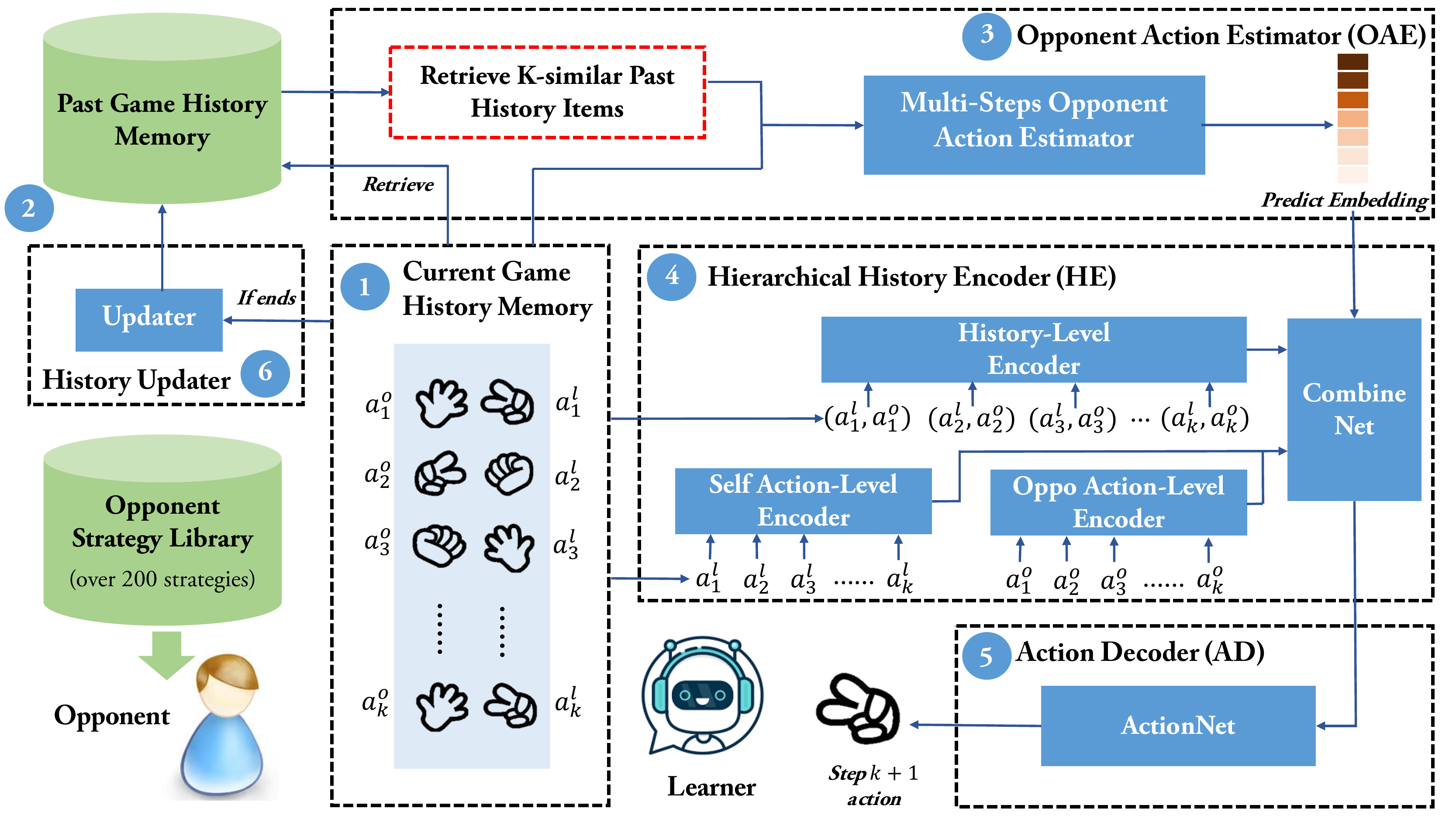} 
\caption{The diagram of our framework that consists of six main components: (1) Current Game History, (2) Past History Memory, (3) Opponent Action Estimator (OAE), (4) Hierarchical History Encoder (HE), (5) Action Decoder (AD) and (6) History Updater (HU).}
\label{framework}
\end{figure}

In order to make the agent have the fast adaptability, we propose a novel method of modeling \emph{Past game history} and a novel framework (named, F3) that fuses current and past game history. As shown in Figure~\ref{framework}, the framework consists mainly of six components: (1) \emph{Current Game History Memory}, (2) \emph{Past Game History Memory}, (3) \emph{Opponent Action Estimator} (OAE), (4) \emph{Hierarchical History Encoder} (HE), (5) \emph{Action Decoder} (AD) and (6) \emph{History Updater} (HU).

Given the \emph{Current game history}, the Opponent Action Estimator (OAE) module with a structure akin to memory networks~\cite{ref20} predicts a embedding vector which represents an estimation of the opponent’s future behavior based on \emph{Past game history}. The Hierarchical History Encoder(HE) module will extract information from different dimensions of the game history and fuse the estimated vector and the current game history representation.

We experiment on several classic repeated games whose number is fixed. In each game, the opponent's strategy is different. Because in the real world, the opponents are usually various and the turns of interactions are multiple. The experimental results show that good results can be achieved only by using past history. The highest score difference can be achieved by fusing past and current history. The Opponent Action Estimator (OAE) model can estimate the opponent's next action. It can be reused in new games which means it has the cross-game adaptability.

In summary, our main contributions are as follows:

\begin{itemize}
    \item  In this paper, we proposed a framework to model the \emph{Current game history} and the \emph{Past game history} in the meantime. 
    \item The model and framework we proposed can make learners to be fast adaptive agents. They can quickly defeat opponents on new strategies. 
    \item The OAE model can be reused in new games. So this framework is easy to implement quickly even in the face of new games, which makes agents have the cross-game adaptability.
\end{itemize}

\section{Method}

\subsection{Basic Notation}

To simplify exposition, we start with some notations.  We use $p_{l}$ to represent the learner (our trained agent). The set of opponents players is denoted by $\mathcal{P}=\left\{ p_{1}, p_{2}, ..., p_{n} \right\}$. $\mathcal{P}$ is consistent with the \emph{Opponent Strategy Library}. Each opponent $p_{i}$ represents a strategy. The action set denoted by $\mathcal{A}=\left\{ a_{1}, a_{2}, ..., a_{n} \right\}$.
The set action of the learner is the same as the opponent's. A complete repeated game $G$ consists of $n$ turns stage games $g_{i}$. Thus, $G=\left\{ g_{1}, g_{2}, ..., g_{n} \right\}$.

Before each repeated game $G_{i}$ starts, the agent $p_{l}$ randomly selects an opponent $p_{o}$ from the opponents set $\mathcal{P}$, then they $(p_{l},p_{o})$ play $n$ turns stage games. Note that all players cannot determine the upper turns limit $n$, which means they do not know when the $G_{i}$ will end. In every stage game $g_{i}$, the players choose actions from $\mathcal{A}$. A stage game action profile $a^s=(a^l,a^o)$ consists of the action of player $p_{l}$ and the action of the opponent player $p_{o}$, then they will separately get a stage reward $r^l$, $r^o$ from the payoff function $g:\mathcal{A} \Rightarrow \mathbb{R} $. When a complete repeated game $G_{i}$ is over, the final score of each player is the average of all stage rewards:

\begin{small}
\begin{equation}
    R^l = \frac{1}{n} \sum_{i=1}^n r^l_{i},  R^o = \frac{1}{n} \sum_{i=1}^n r^o_{i}
\label{e1}
\end{equation}
\end{small}

When a complete repeated game $G_{i}$ ends, \emph{History Updater} decides whether it will be added to the \emph{\textbf{Past Game History Memory}} $ P=\left\{ G_{1}, G_{2}, ..., G_{m} \right\}$. If a repeated game $G_{i}$ proceeds to step $k$, the set of $k-1$ step actions before is defined as \emph{\textbf{Current Game History Memory}} $ C=\left\{ a^s_{1}, a^s_{2}, ..., a^s_{i}...,a^s_{k} \right\}$, $a^s_{i}=(a^l_{i},a^o_{i})$. Thus the entire history $H = P + C$.

\subsection{Framework Overview}

The framework is shown in Figure~\ref{framework}. The \emph{Current Game History Memory} is the actions that have been made in the current game which includes yours and your opponent's. (We assume that actions by both parties can be observed.)  The \emph{Past Game History Memory} stores some complete history which have terminals and final scores. The \emph{Opponent Action Estimator} inputs the current history and find K-similar past history items and outputs a predict vector about the opponent's future behavior through a network. It contains one step opponent action estimation and multiple steps opponent action estimation.  The \emph{Hierarchical History Encoder} fuses current history and the predict vector. The \emph{Action Decoder} converts vectors into the final action.If the current game is over and gets the final scores, the \emph{History Updater} will update the \emph{Past History Memory}.

\subsection{Opponent Action Estimator (OAE)}

Many new strategies are integrations or variants of the old. The past history contains more knowledge to defeat new opponents. But facing new challenges, how to find entries similar to the current problem in \emph{Past game history} and extract useful information from them is a huge challenge. We propose \emph{Opponent Action Estimator} (OAE) to solve this problem.

\begin{figure}
\centering 
\includegraphics[width=10.5cm]{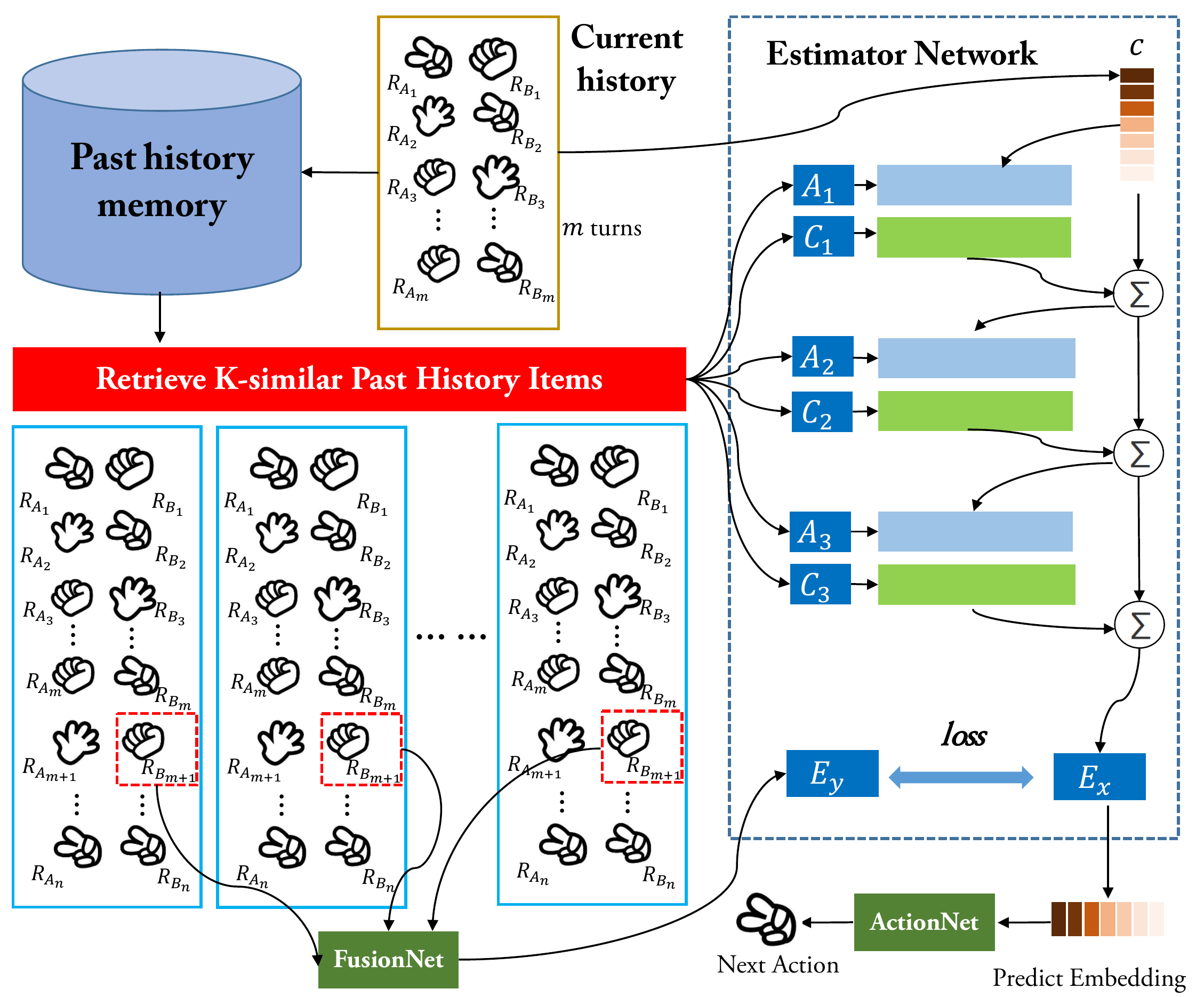} 
\caption{The One-Step Opponent Action Estimator (O-OAE)} 
\label{oae}
\end{figure}

Given the \emph{Current game history} $C$ of $m$ turns, the OAE module predicts a embedding vector which represents an estimate of the opponent's future behavior. This estimation is based on \emph{Past game history}.

First , the $m$ turns $C = \left\{ (a^l_{1},a^o_{1}), (a^l_{2},a^o_{2}), ..., (a^l_{m},a^o_{m}) \right\}$ is used to find the K-similar items from the past game history. As we defined in Section 3.1, the \emph{Past History Memory} $ P=\left\{ G_{1}, G_{2}, G_{3}, ..., G_{t} \right\}$. In $P$, each game $G_{i} = \left\{ (a^l_{1},a^o_{1}), (a^l_{2},a^o_{2}), ..., (a^l_{m},a^o_{m}), (a^l_{m+1},a^o_{m+1}),...,(a^l_{n},a^o_{n}) \right\}$. Each $G_{i}$ will be divided into two subsets: $G^m_{i} = \left\{ (a^l_{1},a^o_{1}), (a^l_{2},a^o_{2}), (a^l_{3},a^o_{3}),..., (a^l_{m},a^o_{m}) \right\}$ and $G^{n-m}_{i} = \left\{ (a^l_{m+1},a^o_{m+1}),...,(a^l_{n},a^o_{n}) \right\}$ .We calculate the similarity betwwen $C$ and each $G^m_{i}$ to find the K-similar items $ G^{sim}=\left\{ G_{1}, G_{2}, ..., G_{k} \right\}$. 

Second, the estimate network, a deep neural network with a structure akin to memory networks~\cite{ref20}, inputs the $m$ turns \emph{Current game history} $C$ and the K-similar items $ G^{sim}$ and outputs an embedding vector $E_{x}$. The \emph{Current game history} $C$ is embedded into a d-dimensional representation $c$ using LSTM. At each layer $l$ of the model, the K-similar items $G^{sim}$ was also embedded into $A_{i}$ and $C_{i}$ using two LSTMs. The weight tying of the estimate network, we
take \textbf{Adjacent} : the output embedding for one layer is the input embedding for the one above, i.e. $A_{k+1} = C_{k}$.~\cite{ref23} The output of each layer is calculated by adding the weighted sum of another embedding of the similar items to he previous item representation. 

Finally, from the K-similar items, based its $G^{n-m}_{i}$, we extract the opponent's action set $O_{i}$ after the step $m$. $O_{i}=\left\{a^o_{m+1},a^o_{m+2},...,,a^o_{n}\right\}$. The opponent's action set of the K-similar items $O=\left\{O_{1},O_{2},...,O_{k}\right\}$ is also fed into a network, FusionNet, to generate a vector $E_{y}$. The structure of the FusionNet is LSTM~\cite{ref18}, followed by a Multi-Layer Perceptrons (MLP).

Our goal is to make $E_{x}$ and $E_{y}$ equal. So we learn a estimate network $\phi_{1}$ parame-terised by $\alpha_{1}$ and a fusion network $\phi_{2}$ parame-terised by $\alpha_{2}$ by minimising the following loss:

\begin{small}
\begin{equation}
    \ell_{OAE}  = \left| \phi_{1}(C, G^{sim}; \alpha_{1}) - \phi_{2}(O;\alpha_{2}) \right| =  \left|  E_{x} -  E_{y} \right|
\label{e2}
\end{equation}
\end{small}

\paragraph{One-Step Opponent Action Estimator (O-OAE)} From each item $G^{sim}_{i}$, if we only extract the opponent's action of step $m+1$, then $O_{i}=\left\{a^o_{m+1}\right\}$. This method is called One-Step Opponent Action Estimator (see Figure~\ref{oae} for details).

\paragraph{Multi-Steps Opponent Action Estimator (M-OAE)} From each item $G^{sim}_{i}$, if we extract all the opponent's action from $G^{n-m}_{i}$, then $O_{i}=\left\{a^o_{m+1},...,,a^o_{n}\right\}$. This method is called Multi-Step Opponent Action Estimator.

\subsection{Hierarchical History Encoder (HE)}

Game is an interactive process. Sometimes different strategies can only be manifested according to the behavior of the opponent. So we use \emph{Hierarchical History Encoder} (HE) to extract information from different dimensions of the game history.

In repeated games, given the current game history $C = \left\{ (a^l_{1},a^o_{1}), (a^l_{2},a^o_{2}), ..., (a^l_{n},a^o_{n}) \right\}$, we use \emph{Action-Level Encoder} and \emph{History-Level Encoder} to encode it. If the length of the game history $C$ is $n$, the \emph{Action-Level Encoder} is a LSTM $f^{action}: \mathbb{L}^{n \times 1} \Rightarrow	\mathbb{L}^{d}$ mapping the agent's action $a_{self} = ( a^l_{1}, a^l_{2}, ..., a^l_{n})$  or the opponent's action $a_{oppo} = (a^o_{1}, a^o_{2}, ..., a^o_{n})$  into a $d$-dimensional vector $h^{action}_{self}$ or $h^{action}_{oppo}$ as the action-level representation of the current history; and the \emph{History-Level Encoder} is a LSTM $f^{history}: \mathbb{L}^{n \times 2} \Rightarrow	\mathbb{L}^{d}$ mapping the union action $C = \left\{ (a^l_{1},a^o_{1}), (a^l_{2},a^o_{2}), ..., (a^l_{n},a^o_{n}) \right\}$  into a $d$-dimensional vector $h^{history}$ as the history-level representation. So there are two \emph{Action-Level Encoder}s and one \emph{History-Level Encoder} in our games. If the number of players is $t$, you can use the $t$ \emph{Action-Level Encoder}s to encode each player's action and then the \emph{History-Level Encoder} wiil be a LSTM $f^{history} : \mathbb{L}^{n \times t} \Rightarrow	\mathbb{L}^{d}$.

Through OAE, we will get an embedding vector $E_{x}$ which represents an estimate of the opponent's future behavior. From HE, we can get $h^{action}_{self}$, $h^{action}_{oppo}$ and $h^{history}$ which  represents the feature of the current history. Then, we use \emph{CombineNet}, a MLP, to fuse these vectors and get a $d$-dimensional vector $h_{c}$.

\subsection{Action Decoder (AD)}

Through the above module, we can get $h_{c}$. The last step is to turn this vector into the next action for the learner. We use \emph{Action Decoder} (AD) to get the final action. The size of the action set $\mathcal{A}=\left\{ a_{1}, a_{2}, ..., a_{s} \right\}$ is $s$. $h_{c}$ is a $d$-dimensional vector. We can also use a MLP $f^{decoder}: \mathbb{M}^{d} \Rightarrow	\mathbb{M}^{s}$ to map the $h_{c}$ to a $s$-dimensional one-hot vector $a$.

If you only want to encode \emph{Past game history}, you can also get the final action directly from OAE. The OAE outputs a $s$-dimensional embedding vector $E_{x}$ which represents an estimate of the opponent's future behavior, then $a = $MLP$(E_{x})$ (see Figure~\ref{oae} for details).

\subsection{Past History Updater (HU)}

When the players receive the end signal, the current game $G_{i}$ is over. Then the agent will get the final reward $R^{l}$ and the opponent will get the final reward $R^{o}$. As shown in formula (1), the final score of each player is the average of all stage rewards  from the payoff function $g$. The score difference $\Delta R = R^{l} - R^{o}$. The action set $G=\left\{(a^l_{1},a^o_{1}), (a^l_{2},a^o_{2}), ..., (a^l_{n},a^o_{n}) \right\}$, the final reward $R^{l}$ and $R^{o}$ compose a complete history item $h = \left\{G, R^{l}, R^{o} \right\}$. 

When the framework starts, if the \emph{Past History Memory} is empty or the maximum capacity is not reached, history item $h$ will be added to the memory. If the memory reaches its maximum capacity, when the current game is over, history item $h$ will be added to the memory and meanwhile the item with the smallest difference will be deleted.

Note that once the OAE and \emph{Past History Updater} are trained, their parameters will be fixed. When the current game is over, we use the final score difference $\Delta R = \overline{R^{l}} - \overline{R^{o}}$ as a guide and use REINFORCE algorithm~\cite{ref19} to update the networks parameters which contain \emph{Hierarchical History Encoder} and \emph{Action Decoder}.

\section{Experiments}

\subsection{Game Setting}

We experiment on two classic games: \emph{Prisoner’s Dilemma} and \emph{Chicken}. They have different reward functions. Their payoff functions are shown in Table~\ref{tab:tb1} and Table~\ref{tab:tb2}. Axelrod\footnote{\url{http://axelrod.readthedocs.org/}} is a python library for the Iterated Prisoner's Dilemma which can enable the reproduction of previous Iterated Prisoner's Dilemma research as easily as possible and create the de-facto tool for future Iterated Prisoner's Dilemma research. It has access to over 200 strategies, including original and classics like \emph{Tit For Tat} and \emph{Win Stay Lose Shift}. New strategies will be encouraged to contribute and new games are easy to defined. Note that all players cannot determine the upper turns limit $n$ in finitely repeated games, which means they do not know when the $G_{i}$ will end. In our experiments, the game turns is fixed. Each game will play 50 turns.


\begin{table*}
  \begin{floatrow}
    \capbtabbox{
    \begin{tabular}{c|cc}
     \toprule
                & Cooperate(C) & Defect(D)\\
    \hline
    Cooperate(C) & 3,~3  & 0,~5  \\
    Defect(D) & 5,~0  & 1,~1   \\
    \toprule
   \end{tabular}
  }
  { \caption{\emph{Prisoner’s Dilemma} Payoff.}
    \label{tab:tb1} }
 \capbtabbox{
   \begin{tabular}{c|cc}
   \toprule
      & Swerve(S) & Going(G) \\
   \hline
    Swerve(S) & 2,~2  & 1,~5 \\
    Going(G) & 5,~1  & 0,~0  \\
   \toprule
   \end{tabular}
  }
  { \caption{\emph{Chicken} Payoff.}
    \label{tab:tb2}
   }
  \end{floatrow}
\end{table*}

\subsection{Evaluation Metric}
 When the current game is over, the agent will get the final reward $R^{l}$ and the opponent will get the final reward $R^{o}$. In multiple tests, we calculate their average scores  $\overline{R^{l}}$ and $\overline{R^{o}}$ (see Equation~\ref{e1}).In all games, the number of turns is fixed. Therefore, a player with a higher score means mastering the opponent ’s strategy more quickly, which means it has faster adaptability. The difference $\Delta R = \overline{R^{l}} - \overline{R^{o}}$ is the evaluation metric.

\subsection{Baselines}


\paragraph{Q-Learning}  Q-learning~\cite{ref21} is the easiest way to use \emph{Current game history}.  In this paper, experiments are performed on Second-order Markov process.

\paragraph{Deep Q Network} Deep Q learning (DQN)~\cite{ref22} leverages advances in deep learning to learn policies from high dimensional sensory input. Different from Atari 2600 games using convolutional networks, we use Long Short Term Memory (LSTM) to convert the game history to vectors. The \emph{Current game history} $ C=\left\{ a^s_{1}, a^s_{2}, ..., a^s_{i}...,a^s_{k} \right\}$ , so $ h = $LSTM$(C)$. A two-layer Multi-Layer Perceptrons (MLP) is then used to predict the next action. 

\paragraph{Policy Gradient} The network structure is the same as DQN which uses LSTM to encode the current history and uses MLP to map the action set $\mathcal{A}$. When the current history ends, we train the networks using Policy Gradient~\cite{ref19}. The network will only be updated when the game is over.

\subsection{Use Current History Only}

\begin{table*}[t]
\centering
\begin{tabular}{l|cc|cc}  
\toprule
\quad & \multicolumn{2}{c|}{Prisoner’s Dilemma} & \multicolumn{2}{c}{Chicken}  \\
\hline
 Model & Old Strategies & New Strategies  & Old Strategies  & New Strategies  \\
\hline
Q-learning  & -1.022 &  -1.055 & -0.872 & -0.840  \\
HE+AD(DQN)     & 0.522 &  0.301 & 0.434 & 0.461  \\
HE+AD(PG)   & 1.199 &  1.091 & 1.105 & 0.943   \\
\hline
O-OAE+AD  & 1.148 &  0.958 & 0.842 & 0.764  \\
M-OAE+AD  & 1.079 &  0.894 & 0.851 & 0.793  \\
\hline
O-OAE+HE+AD    & 1.288 &  1.116 & 1.123 & 1.009  \\
M-OAE+HE+AD    & 1.236 &  1.118 & 1.131 & 1.001  \\
\bottomrule
\end{tabular}
\caption{The average scores difference $\Delta R = \overline{R^{l}} - \overline{R^{o}}$. The \emph{Old Strategies} is the training strategies set and \emph{New Strategies} is the new strategies set. After each 500 times training epoch, we will test one time.}
\label{re1}
\end{table*}

\begin{table*}
\centering
\begin{tabular}{l|cc|cc}  
\toprule
\quad & \multicolumn{2}{c|}{New Trained OAE} & \multicolumn{2}{c}{Reused OAE}\\
\hline
 Model & Old Strategies & New Strategies  & Old Strategies  & New Strategies  \\
\hline
O-OAE+AD  & 0.840 &  0.747 &  0.842 & 0.764  \\
M-OAE+AD  & 0.854 &  0.780 &  0.851 & 0.793  \\
\hline
O-OAE+HE+AD    & 1.134 &  1.010 & 1.123 & 1.009  \\
M-OAE+HE+AD    & 1.139 &  1.031 & 1.131 & 1.001  \\
\bottomrule
\end{tabular}
\caption{The average scores difference $\Delta R$ of \emph{Reused OAE}.}
\label{re2}
\end{table*}

Although after each stage game, all players will receive a stage reward, but in the finite repeated game, the final reward is the basis for determining the outcome when the game is over. Q-Learning only relies on the history of the previous n turns to make the next decision, instead of making decisions from the whole of the finite repeated game. DQN relies on all current history, but it only considers the next decision, which is obviously greedy. Because sometimes, in order to maximize the overall benefit, players may make compromises in some stage games. While, the Policy Gradient methods target at modeling and optimizing the policy directly. In a finite repeated game, the opponent's strategy usually constant. So the Policy Gradient can model this strategy and defeat it.

From the Table~\ref{re1} we can see that the Q-Learning score is negative, which means that the method can not beat the opponent at all. The DQN method beats its opponent by about 0.5 points on the \emph{Old Strategies} and about 0.3 points on the \emph{New Strategies}. While the Policy Gradient score difference exceeds 1 whether it is a new strategy or an old strategy. Experiments show that the Policy Gradient is suitable for finite repeated game. Note that the first three rows of experiments in Table~\ref{re1} only use the \emph{Current game history}. Since Policy Gradient is the best method for finite repeated games, all next experiments use this method for optimization.

\subsection{The Performance of OAE}

The Opponent Action Estimator(OAE) module use current history to retrieve the K-similar items form the past history and generates an estimated vector. To test the performance of OAE, after getting the estimated vector about the opponent's next step, we use Action Decoder (AD)  directly to decode it into the action. We use the Policy Gradient~\cite{ref19} to train the AD module. As shown in Figure~\ref{re1},  although the scores do not reach the level of using only the current history, but also achieved a good score. 

\subsection{Comparative Experiments}

The current history is the most important basis for decision-making, we use Hierarchical Encoder to encode it. The optimal response of strategies is stored in the past game history, it can estimate the opponent's next action and provide more information for the best response. As shown in Figure~\ref{re1}, the highest score difference can be achieved by fusing \emph{Past game history} and \emph{Current game history}.

The One-Step Opponent Action Estimator (O-OAE) only relies on the opponent's one-step decision in past history while the Multi-Steps Opponent Action Estimator (M-OAE) use opponent's multi-steps decision. But as shown in Figure~\ref{re1}, whether it is directly using OAE or fusing with the past history, we find that they have no obvious difference.

\subsection{Reuse of OAE}

Given the Current game history, the OAE module akin to memory networks~\cite{ref20} predicts a embedding vector which represents an estimate of the opponent’s future behavior. Essentially, it is a retrieval model. As long as the current history and past history belong to the same game, the OAE can be reused.  We trained the O-OAE and M-OAE on \emph{Prisoner’s Dilemma}, but we reuse them on \emph{Chicken Game}. We experiment two kinds of OAE on \emph{Chicken Game}: (a) the \emph{New Trained OAE} is the Opponent Action Estimator module trained on the \emph{Chicken} game and tested on the \emph{Chicken} game and (b) the \emph{Reused OAE} is the module trained on the \emph{Prisoner’s Dilemma} game and reused on the \emph{Chicken} game. 

\section{Related Work}

When a static non-cooperative strategic game is repeated over time, players iteratively interact by playing a similar stage game. This type of game is called as a repeated game. It consists of a number of repetitions of some base game (also called a stage game)~\cite{ref8}. Repeated games may be broadly divided into two classes, finite and infinite, depending on how long the game is being played for. Infinite games are those in which the game is being played an infinite number of times. Finite games are those in which both players know that the game is being played a specific and finite number of rounds. When the game ends, all the players will receive end terminal~\cite{ref9} and final rewards.

In the fields of \emph{Economics}, related research has focused on Modeling Bounded Rationality.~\cite{ref9} proposed the set of histories $H$ can be partitioned into three subsets: (1) $Z$: the set of terminal histories; (2) $C$: the set of histories after which the chance player moves and (3) $D=H-Z-C$: the set of histories after which the decision maker moves. Inspired by this, we define the set of terminal histories $Z$ the \emph{Past history} and explore the fusion of \emph{Past game history} and \emph{Current game history}.

Current studies about the use of historical information and limited memory in repeated games are mainly limited to the \emph{Current game history}. In Multi-Agent Reinforcement Learning (MARL), ~\cite{ref17} proposed a new algorithm which demonstrates strong performance in empirical tests against a variety of opponents in a wide range of environments. But the bounded memory here is limited to \emph{Past history}.~\cite{ref18} talked about the Memory-one strategies. This manuscript investigates best responses to a collection of memory-one strategies as a multidimensional optimisation problem. But in this article, we not only discuss the current history, but also try to use the \emph{Past history}.

In the research of \emph{Game Theory}, the new strategies on the prisoner's dilemma and other games are constantly being proposed: \emph{Tit For Tat}~\cite{ref16}, \emph{Spiteful Tit For Tat} (Prison1998), \emph{Hard Tit For Two Tats}~\cite{ref14}, \emph{Suspicious Tit For Tat} ~\cite{ref15} and so on. Our proposed method can defeat the new strategy based on learning the old strategy, which makes agents easily cope with the new strategies that are constantly proposed.

The method of modeling teammates in MARL also can quickly defeat the opponents. ~\cite{ref12} proposed Self Other-Modeling (SOM), in which an agent uses its own policy to predict the other agent’s actions and update its belief of their hidden state in an online manner.~\cite{ref13} presented Learning with Opponent-Learning Awareness (LOLA), a method that reasons about the anticipated learning of the other agents. They treat all their opponents as one type. But in the real world, the type of opponents may be variable. In a new game, we may meet opponents that we have never encountered before.

\section{Conclusion}

In this paper, We propose a model to make use of past history and a framework to fuse past history and current history. The model and framework we proposed can make learners to be fast adaptive agents. They can quickly defeat opponents on new strategies.The experimental results show that good results can be achieved only by using past history. The highest score difference can be achieved by fusing past and current history. The OAE model can be reused in new games which means it has the cross-game adaptability.

\section*{Broader Impact}


Whether it is a dialogue or a game, in the process of human interaction, it needs an agent with fast adaptability.The model and framework proposed in this paper can make better use of past and current game history, so that the agent can adapt quickly.

The agent has the ability to adapt quickly, and can bring a better experience to humans in human-computer interaction. AI can be used to develop more agents, which makes AI serve humans better.

\end{document}